\documentclass[a4paper,12pt]{article}
\usepackage[top=2cm, bottom=2cm, left=2.5cm, right=2.5cm]{geometry}
\usepackage[utf8]{inputenc}
\usepackage{amsmath, amsfonts, amssymb}
\usepackage[english]{babel}
\usepackage{indentfirst}
\usepackage[T1]{fontenc}
\usepackage{authblk}
\usepackage{graphicx} 
\usepackage{subfigure} 
\usepackage{dcolumn}
\usepackage{bm}
\usepackage{color}
\newcommand{\integral}{\displaystyle\int} 
\newcommand{\somatorio}{\displaystyle\sum} 
\date{}

\author[1]{Ivandson Praeiro de Sousa}
	\affil[1]{\small Departamento de Física Teórica e Experimental, Universidade Federal do Rio Grande do Norte, Av. Sen. Salgado Filho, 3000
- Lagoa Nova, CEP. 59078-970, Natal, RN, Brasil
		}

\begin{document}

\title{Phase Transition on the Classical Dimer Model on a 3-4 Lattice\\}

\maketitle

	\begin{abstract}
In this paper, we investigate if it occurs a phase transition for a system of classical dimers adsorbed on a 3-4 lattice in the thermodynamic limit. We define four types of bonds (denoted by the dimers activities adsorbed in her, $x$, $y$, $u$ and $v$) and evaluated the partition function from the combinatorial method of the pfaffian and using periodic boundary conditions. We write the density of free energy per particle and the condition to his analyticity, showing that there is a critical point for a determined choice of the activities. We also numerically evaluated the integral on free energy and their derivates (entropy and specific heat) for the same case and compute the entropy density and molecular freedom per particle at high temperatures for this choice of the activities.\\

	\end{abstract}

\section{Introduction}

The dimer model has been proposed in 1937 to explain the thermodynamic behavior of a gas of diatomic molecules (dimers) adsorbed\footnote{The term \textit{adsorption} are associated to a superficial process. It refers to the phenomena which a gas stays on a surface of a crystalline material. This term does not have to be confounded with \textit{absorption}, which do not is a surface phenomenon.} on a crystalline surface \cite{fowler}. In the 1960s emerged solutions based on combinatorial methods, in which were obtained the partition function for the system of dimers in bidimensional lattices \cite{kasteleyn61, fisher61, fisher61_2, temperley61}. In fact, the harder part in the solution of this model consists to determinate the number of ways to arrange an ensemble of dimers in a lattice.

The dimer model is one of the few nontrivial lattice models in statistical mechanics that may be exactly solved, and until today arouses the interest of mathematics and physicists \cite{li, giuliani, aizenman, casagrande, wu2016}. 

The existence of a phase transition in dimer models depends on the geometry of the lattice and on the choice of the bond activities \cite{casagrande}. A search in the literature indicates that still does not exist treatment of dimer model in a 3-4 lattice. So, this paper does it.

In this paper, we investigated if it occurs a phase transition in dimer model in a 3-4 lattice with periodic boundary conditions. For this, we use the combinatorial method of the pfaffian developed by Kasteleyn \cite{kasteleyn63}.

We begin the text with an introduction to method together with some examples of his application in planar lattices. After, we show the results obtained for 3-4 lattice, followed by his discussion.

\section{Method}

In this section, we present the combinatorial method that we use to determinate the partition function for the dimers in a generic planar lattice. We consider in all this text only configurations known as the closed-packed dimer model, where the lattice is entirely covered with dimers, without there are holes or isolated particles (monomers) \cite{nagle-yokoi}. This configuration is known in the graph theory as \textit{perfect matching} \cite{kenyon}.

\begin{figure}[h!]
	\centering
	\includegraphics[scale=1.0]{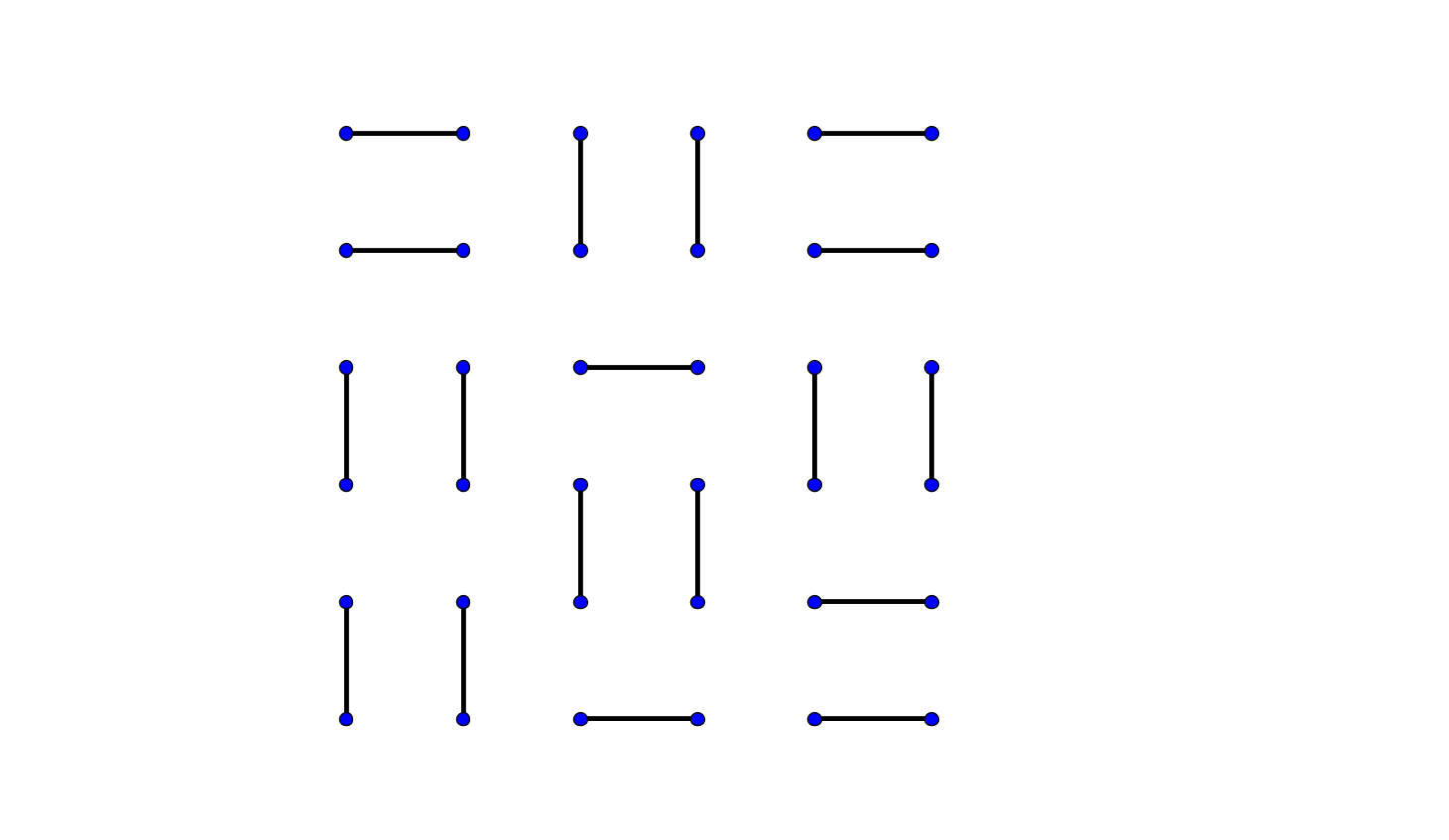}
	\caption{An example of dimer configuration in a quadratic lattice.} 
	\label{quadrada} 
\end{figure}

The problem of statistical of closed-packed dimers on a planar lattice may be summarised in the following way \cite{kasteleyn61, fisher61, temperley61, nagle-yokoi, kenyon}: \textit{given a lattice with $N$ sites, how to cover completely the lattice with dimers without there are holes?} That number naturally it depends to lattice shape. For example, we show in the figure \ref{quadrada} a dimer configuration possible for the quadratic lattice. In other words, the figure \ref{quadrada} shows a perfect matching possible in a quadratic lattice with $6 \times 6$ sites\footnote{Note that this is just a manner of to arrange the dimers in a way that the condition that not there are holes is satisfied. For the partition function calculus, we need determinate the number of possible dimer configurations, which named multiplicity of the dimer states.}.

The total energy of the dimers adsorbed on a two-dimensional lattice, in which the dimer that stays in a given bond of the lattice have a $\epsilon_b$ energy (for example, horizontal bonds, vertical bonds) \cite{kasteleyn61, fisher61, temperley61, nagle-yokoi, kenyon}, is given by

\begin{equation}
	E = \somatorio_b \epsilon_b.
\end{equation}

\noindent So, the partition function for dimers adsorbed on a lattice with several types of bond b is

\begin{equation}
	Z(\{N_b\}) = \somatorio_C g_C(\{N_b\}) \ \exp{\left(-\beta \somatorio_b \epsilon_b \right)} = \somatorio_C g_C(\{N_b\}) \prod_b z_b^{N_b} ,
\end{equation}

\noindent in that $g_C$ denote the dimers states multiplicity \cite{kasteleyn61, fisher61, temperley61, kenyon}. In other words, $g_C$ denote the number of perfect matchings for a given lattice. The $z_b$ term (the Boltzmann factor) is the activity of the dimers in the $b$ bond ($z_b = e^{- \beta \epsilon_b}$) and $N_b$ denotes the number of bonds of $b$ type. The basic question to solve any dimer problem is how determinate $g_C$.

The answer to this question was provided by Kasteleyn, which is known as the pfaffian method \cite{kasteleyn61, kasteleyn63, nagle-yokoi, montroll}. Whit his theorem, which has been famous in graph theory \cite{kenyon}, kasteleyn demonstrate that if we associate a special orientation (namely, a pfaffian orientation)  to the graph representing the lattice, the number of dimer configurations ($g_C$) is given by the pfaffian of the adjacency\footnote{The adjacency matrix is a matrix which the entries ($a_{i, j}$) denote the number of bonds connecting the sites $i$ and $j$.} matrix associated to graph. The pfaffian is a polynomial in the entries of the adjacency matrix, which is very similar to the determinant of the matrix. In case of the dimers adsorbed in a given lattice, the entries of the adjacency matrix are 0 or 1 and the pffafian perfectly count the number of possible perfect matchings in the lattice \cite{kasteleyn61, fisher61, temperley61, nagle-yokoi, kenyon, montroll}. The pfaffian of the adjacency matrix is equal to square root of her determinant:

\begin{equation}
	pf(\bm{A}) = \sqrt{det(\bm{A})}.
\end{equation}

The Kasteleyn's theorem establishes that all planar graph has at least one pfaffian orientation. A pfaffian orientation is that in which the bonds are oriented in a way that the number of bonds oriented in the clockwise in any elementary polygon in the lattice is odd \cite{kasteleyn61, kasteleyn63, nagle-yokoi, kenyon, montroll}.

In the thermodynamic limit, the size of lattice goes to infinity, in a manner that the order of matrix is also infinity. However, when we use periodic boundary conditions, the adjacency matrix has a periodic structure, which may be diagonalized in blocs, in a way that it only takes compute the determinant of a matrix which usually only reaches the order 10, since all the blocs are equal \cite{montroll}. In this case, the free energy per particle to the adsorbed dimers on a two-dimensional lattice is given by

\begin{equation}
	\label{receita}
	-\beta f = \dfrac{1}{8\pi^2} \integral_0^{2\pi} d\theta \integral_0^{2\pi} d\phi \ln{\det \bm{\Lambda}},
\end{equation}

\noindent in that $\bm{\Lambda}$ denote the blocs of the adjacency matrix, and they are may be written as \cite{montroll}

\begin{equation}
	\bm{\Lambda} (\theta, \phi) = \somatorio_u \bm{M} (\vec{u}) e^{i\vec{u} \cdot \vec{\psi}},
\end{equation}

\noindent in that $\vec{u} = (u_i, u_j)$ is a vector that identify the position of each unitary cell in the plan and $\vec{\psi} = (\theta, \phi)$. The elements of the matrix $\bm{M}$ denotes the connectivity of the sites in the basic cell, that is, $M_{i, j} = 0$ if there is no bond connecting the sites i and j, 1, if there is one bond oriented from i to j, and -1 if there is one bond oriented from j to i. The order of the $\bm{M}$ matrix is equal to the number of sites in each cell, and the number of the matrices $\bm{M}$ required is equal to the central cell plus the number of the first neighbor cells. In next section, we show two examples of how to use the equation \ref{receita} to determinate the density of free energy of the two classical dimer models, in the triangular and 4-8 lattice.

\subsection{Example in the triangular lattice}

\begin{figure}[h!]
	\centering
	\includegraphics[scale=0.35]{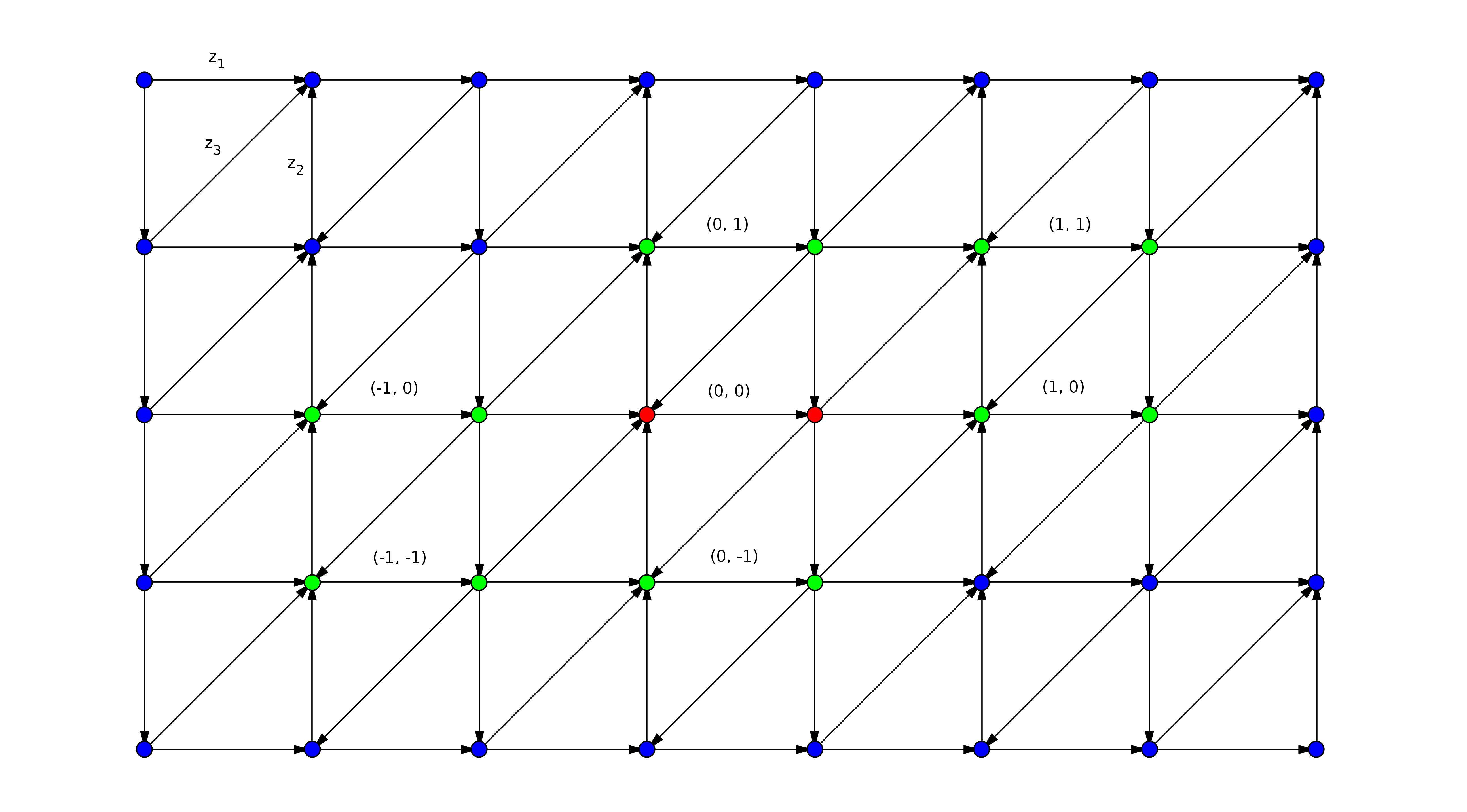}
	\caption{Our scheme for triangular lattice together with the Kasteleyn orientation of edges.} 
	\label{triangular} 
\end{figure}

An interesting dimer problem that presents phase transition considers the dimers adsorbed on the triangular lattice. We show in the figure \ref{triangular} a piece of the triangular lattice together with the basic cells of interest and a pfaffian orientation for the bonds. We show in red the cell that generates all lattice and in green his first neighbor cells.  Now let's apply the method showed in the previous section for this lattice.

As is usually done in papers \cite{wu}, we consider three types of bonds, horizontal, vertical and diagonal. In this case, the matrices $\bm{M}$ associated with unitary cells showed in the figure are

$$
	\bm{M}(0, \, 0) = 
	\begin{bmatrix}
	0 & z_1 \\
	-z_1 & 0
	\end{bmatrix} ;
$$

\

$$
	\bm{M}(-1, \, 0) = 
	\begin{bmatrix}
	0 & -z_1 \\
	0 & 0
	\end{bmatrix} ;
	\bm{M}(1, \, 0) = 
	\begin{bmatrix}
	0 & 0 \\
	z_1 & 0
	\end{bmatrix} ;
	\bm{M}(0, \, -1) = 
	\begin{bmatrix}
	-z_2 & 0 \\
	z_3 & z_2
	\end{bmatrix} ;	
$$

\

$$
	\bm{M}(0, \, 1) = 
	\begin{bmatrix}
	z_2 & -z_3 \\
	0 & -z_2
	\end{bmatrix} ;	
	\bm{M}(-1, \, -1) = 
	\begin{bmatrix}
	0 & -z_3 \\
	0 & 0
	\end{bmatrix}
	\bm{M}(1, \, 1) = 
	\begin{bmatrix}
	0 & 0 \\
	z_3 & 0
	\end{bmatrix}.
$$

\

\noindent So, the matrix $\bm{\Lambda}$ required to calculate the partition function is

\begin{equation}
	\bm{\Lambda} = 
	\begin{bmatrix}
	z_2(e^{i\phi} - e^{-i\phi}) & z_1(1 - e^{-i\theta}) - z_3(e^{i\phi} - e^{-i(\theta + \phi)}) \\
	z_1(e^{i\theta} - 1) + z_3(e^{i(\theta + \phi)} + e^{-i\phi}) & -z_2(e^{i\phi} - e^{-i\phi})
	\end{bmatrix},
\end{equation}

\noindent so that

\begin{equation}
	\label{detri}
	\det \bm{\Lambda} = \left[ z_1^2 + z_2^2 + z_3^2 - z_1^2\cos\theta -z_2^2\cos{2\phi} +z_3^2\cos{(\theta + 2\phi)} \right].
\end{equation}

From the form of the determinant, we to may find the critical points for the dimers adsorbed on triangular lattice \cite{wu}. This may be done by looking at a portion of the plane $(\theta, \phi)$, that is, examining the equation \ref{detri} for a certain value of $\theta$ and $\phi$. This provides three possible critical points,

\begin{equation}
	z_i = 0, \, i=1, 2, 3.
\end{equation}

\pagebreak \subsection{Example in the 4-8 lattice}

\begin{figure}[h!]
	\centering
	\includegraphics[scale=0.6]{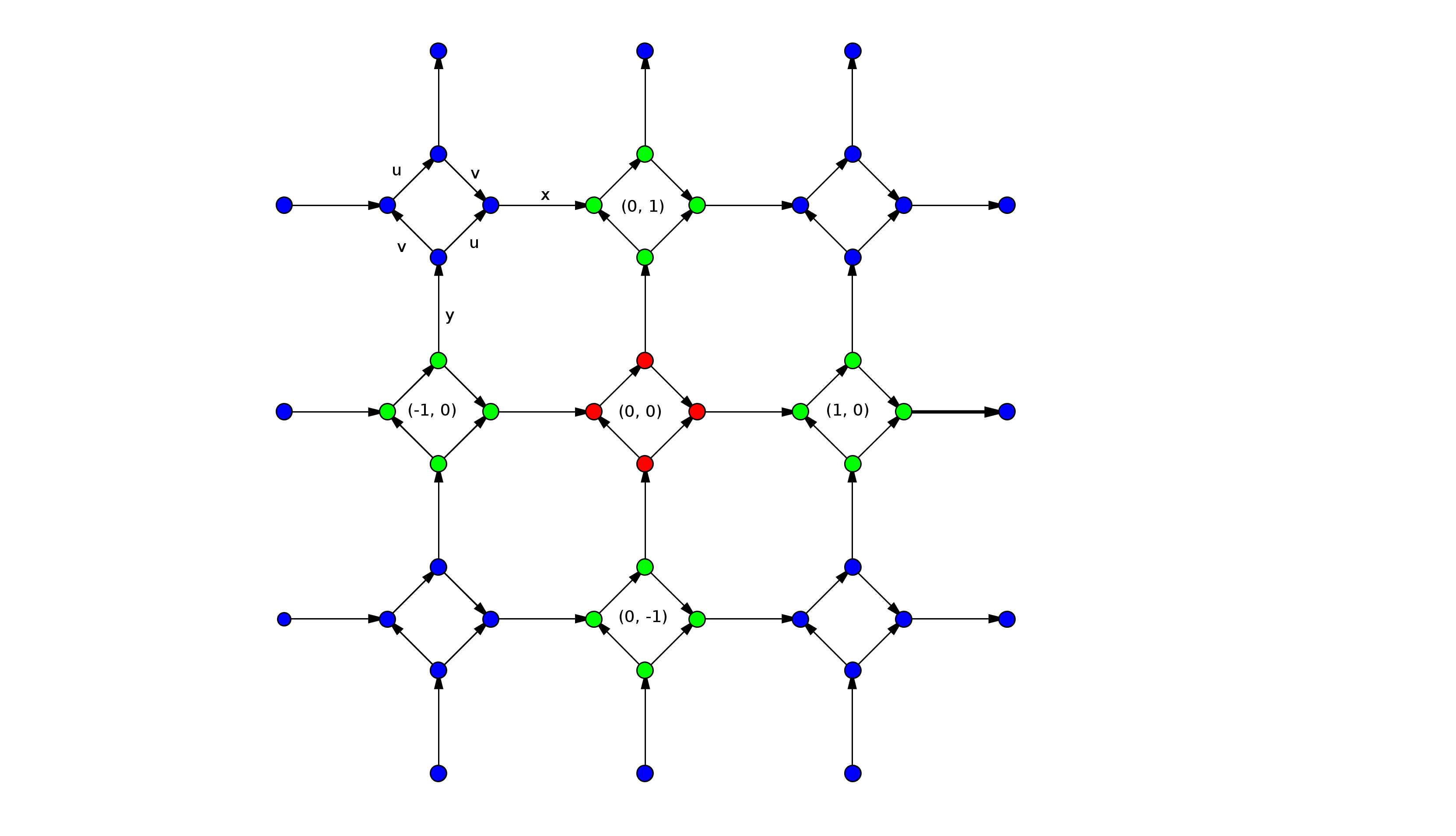}
	\caption{Our scheme for 4-8 lattice together with the Kasteleyn orientation of edges.} 
	\label{48} 
\end{figure}

The dimer model on 4-8 lattice constitutes a problem of fundamental importance since has direct application on phase transition problem of physical systems \cite{casagrande, wu, salinas-nagle}. We show in the figure \ref{48} a peace of the 4-8 lattice and a pfaffian orientation for their bonds.

In this case, we make a distinction between four different types of bounds, that is equivalent to four distinct values of energy. So that, the matrices representing the basic cells are

$$
	\bm{M}(0, \, 0) = 
	\begin{bmatrix}
	0 & v & 0 & -u \\
	-v & 0 & -u & 0 \\
	0 & u & 0 & v \\
	u & 0 & -v & 0 
	\end{bmatrix} ;
$$

\

$$
	\bm{M}(-1, \, 0) = 
	\begin{bmatrix}
	0 & 0 & 0 & 0 \\
	0 & 0 & 0 & 0 \\
	0 & 0 & 0 & 0 \\
	0 & -x & 0 & 0 
	\end{bmatrix} ;
	\bm{M}(1, \, 0) = 
	\begin{bmatrix}
	0 & 0 & 0 & 0 \\
	0 & 0 & 0 & x \\
	0 & 0 & 0 & 0 \\
	0 & 0 & 0 & 0 
	\end{bmatrix} ;	
$$

\

$$
	\bm{M}(0, \, -1) = 
	\begin{bmatrix}
	0 & 0 & 0 & 0 \\
	0 & 0 & 0 & 0 \\
	-y & 0 & 0 & 0 \\
	0 & 0 & 0 & 0
	\end{bmatrix} ;	
	\bm{M}(0, \, 1) = 
	\begin{bmatrix}
	0 & 0 & y & 0 \\
	0 & 0 & 0 & 0 \\
	0 & 0 & 0 & 0 \\
	0 & 0 & 0 & 0
	\end{bmatrix},
$$

\

\noindent in a way that the adjacency matrices blocs have the form

\begin{equation}
	\bm{\Lambda} = 
	\begin{bmatrix}
	0 & v & y \, e^{i\phi} & -u \\
	-v & 0 & -u & x \, e^{i\theta} \\
	-y \, e^{-i\phi} & u & 0 & v \\
	u & -x \, e^{-i\theta} & -v & 0 
	\end{bmatrix}.
\end{equation}

The specific heat of dimers on 4-8 lattice blows up when the fugacities satisfy the relation

\begin{equation}
	\label{critico48}
	xy = u^2 + v^2.
\end{equation}

\section{Results}

\begin{figure}[h!]
	\centering
	\includegraphics[scale=0.35]{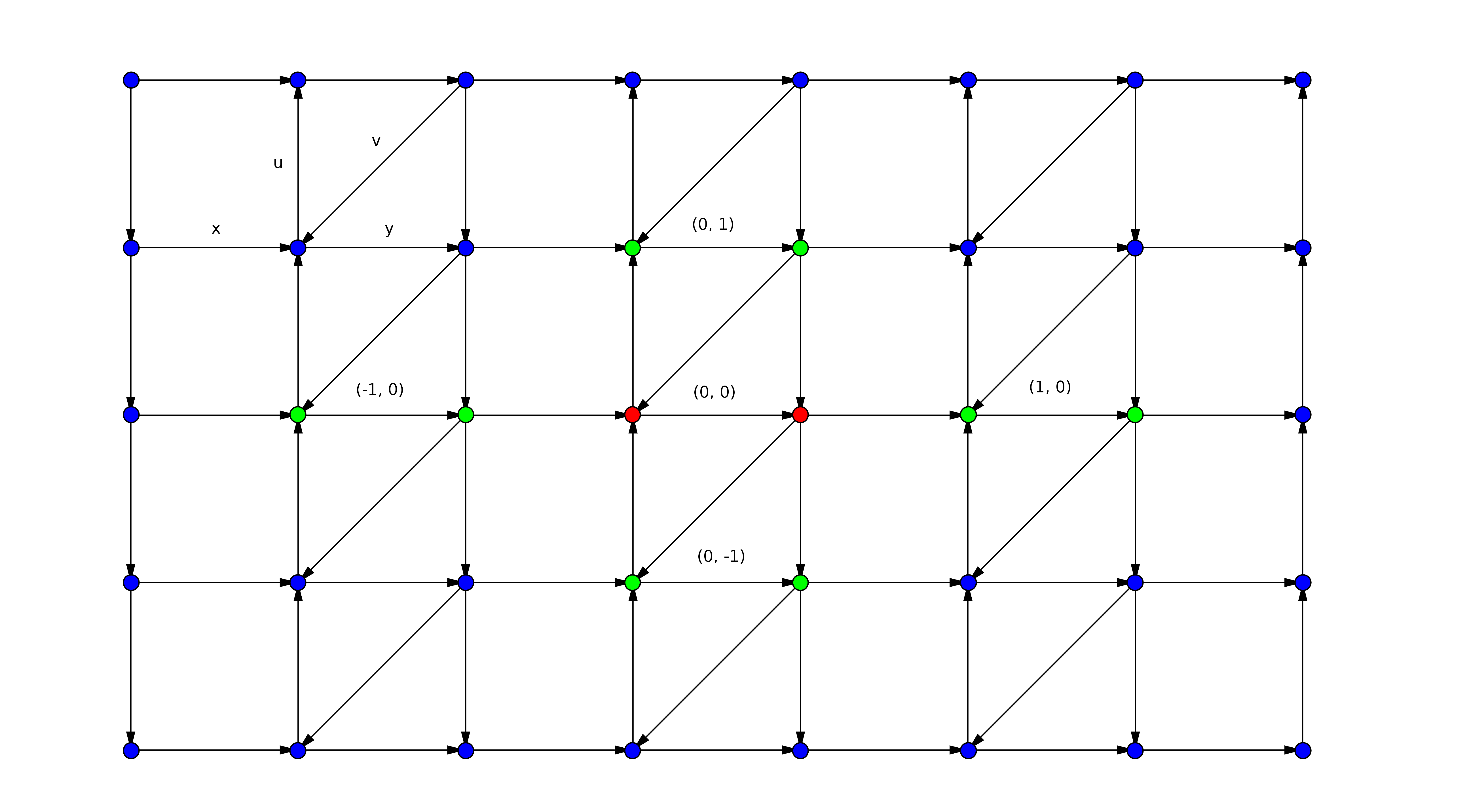}
	\caption{Our scheme for 3-4 lattice together with the Kasteleyn orientation of edges.} 
	\label{341} 
\end{figure}

We show follow our representation for the 3-4 lattice that we consider in this paper (fig. \ref{341}). We make choice four distinct types of the bonds, as is shown in the figure \ref{341}. We also show the Kasteleyn prescription of pfaffian orientation of edges and the relevant basic cells for the pfaffian calculus.

The matrices related to unitary cells are

\begin{eqnarray}
	\bm{M}(0, \, 0) = 
	\begin{bmatrix}
	0 & y \\
	-y & 0
	\end{bmatrix}
\end{eqnarray}

\begin{eqnarray}
	\bm{M}(-1, \, 0) = 
	\begin{bmatrix}
	0 & -x \\
	0 & 0
	\end{bmatrix} ;
	\bm{M}(1, \, 0) = 
	\begin{bmatrix}
	0 & 0 \\
	x & 0
	\end{bmatrix} 
\end{eqnarray}

\begin{eqnarray}
	\bm{M}(0, \, -1) = 
	\begin{bmatrix}
	-u & 0 \\
	v & u
	\end{bmatrix} ;	
	\bm{M}(0, \, 1) = 
	\begin{bmatrix}
	u & -v \\
	0 & -u
	\end{bmatrix} ,
\end{eqnarray}

\noindent so that write the blocs of the adjacency matrix as

\begin{eqnarray}
	\bm{\Lambda} = 
	\begin{bmatrix}
	u(e^{i\phi} - e^{-i\phi}) & y-x \, e^{-i\theta} - v \, e^{i\phi} \\
	-y+x \, e^{i\theta}+v \, e^{-i\phi} & u(e^{-i\phi} - e^{i\phi})
	\end{bmatrix}.
\end{eqnarray}

\noindent We may now write easily the determinant of adjacency matrix and consequently the free energy on thermodynamic limit:

\begin{equation}
	\det \bm{\Lambda} = \left[ 2u^2 + v^2 + x^2 + y^2 -2y(x\cos\theta +v\cos\phi) -2u^2\cos{2\phi} + 2xv\cos{(\theta + \phi)} \right]
\end{equation}

\begin{equation}
	\label{free_energy}
	- \beta f =  \dfrac{1}{4\pi} \integral_0^{2 \pi} d\phi \ln{\left[ u^2 + \frac{1}{2}( v^2 + x^2 + y^2 ) -yv\cos\phi -u^2\cos{2\phi} + \sqrt{q(\phi)} \right]}
\end{equation}

\noindent where

\begin{equation}
	q(\phi) = ( u^2 + \frac{1}{2}( v^2 + x^2 + y^2 ) -yv\cos\phi -u^2\cos{2\phi} )^2 - ( -xy + vx\cos\phi )^2 - v^2x^2\sin^2\phi.
\end{equation}

\noindent The quantity $q(\phi)$ defines the critical point of the lattice, that occur when

\begin{equation}
	v = x + y.
\end{equation}

\noindent If we make $x=y$ and $u=v$, we may simplify the expression in \ref{free_energy}, that may be numerically evaluated in function the value of $z=\frac{y}{v}$, where the critical point is given by $z=\frac{1}{2}$.

\begin{figure}[h!]
		\center
		\subfigure[energia][Free energy per particle.]{\includegraphics[width=7cm]{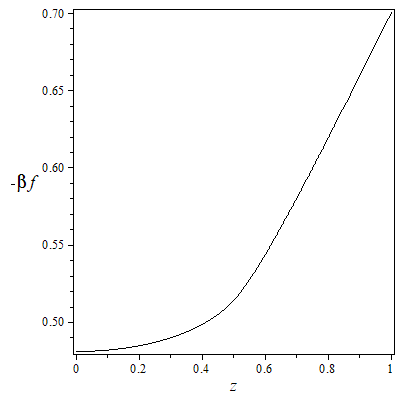}}
		\qquad
		\subfigure[magnetizacao][Entropy per particle.]{\includegraphics[width=7cm]{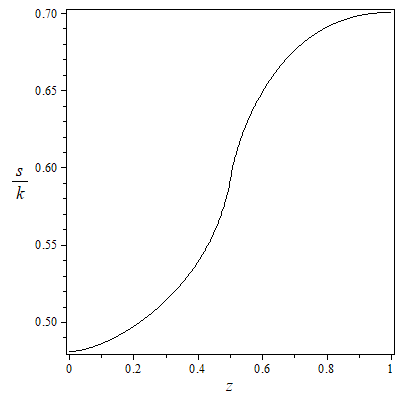}}
		\caption{Plots of free energy and entropy per particle in function of $z$}
		\label{energy_entropy}
	\end{figure}

\begin{figure}[h!]
	\centering
	\includegraphics[scale=0.5]{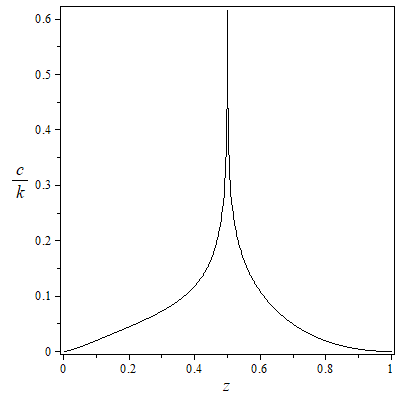}
	\caption{Specific heat per particle in function of $z$.} 
	\label{c_341} 
\end{figure}

We show in figures \ref{energy_entropy} and \ref{c_341} the plots of free energy, entropy and specific heat for particle.

\pagebreak \section{Discussion}

As we say in the previous section, we consider the special case in that $x=y$ and $u=v$, which implies $z = 1/2$ as the critical point. As $z = e^{-\beta \epsilon}$, we may write the critical temperature as 

\begin{equation}
	T_c = \dfrac{\epsilon_2 - \epsilon_1}{k_B \ln 2},
\end{equation}

\noindent in that $\epsilon_1$ and $\epsilon_2$ are given respcetively from $v = e^{- \beta \epsilon_1}$ and $y = e^{- \beta \epsilon_2}$, which is the same expression for the honeycomb lattice, although in other context. 

We numerically integrate the expressions for density of free energy, entropy density and specific heat per dimer (figs. \ref{energy_entropy} and \ref{c_341}). The plots demonstrate that the analytical result suggests, namely, that the specific heat blows up in the critical value of $z$.

A relevant quantity in dimer models is the molecular freedom of adsorbed dimers, which it is identified as the number of accessible configurations per dimer in the limit of the high temperatures,

\begin{equation}
	\bm{\Phi} = \exp{\dfrac{\ln{Z\{z_i=1\}}}{N}} = 2. \, 015 \, 269 \, 613 \, \ldots \ .
\end{equation}

\noindent This quantity is obtained from entropy in high temperatures, which is given by

\begin{equation}
	\dfrac{\ln{\bm{Z}\{z_i=1\}}}{N}=0.700 \, 752 \, 989 \, \ldots 
\end{equation}

\noindent We show in table \ref{quadro} the values of molecular freedom to dimer model in several lattices.

\begin{table}[h!]
	\centering
	\begin{tabular}{c c c}
		\hline
		\textbf{Lattice} & \textbf{distinct bonds} & \textbf{Molecular freedom} \\ \hline
		rectangular & 2 & $1. \, 791 \, 622 \, 812 \, \ldots$ \\ \hline
		honeycomb & 3 & $1. \, 381 \, 356 \, 444 \, \ldots$ \\ \hline
		triangular & 3 & $2. \, 356 \, 527 \, 353 \, \ldots$ \\ \hline
		4-8 & 4 & $1. \, 457 \, 897 \, 968 \, \ldots$ \\ \hline	
		3-4  & 4 & $2. \, 015 \, 269 \, 613 \, \ldots $ \\ \hline
	\end{tabular} 
	\caption{Comparactive frame of lattices molecular freedom.}
	\label{quadro}
\end{table}

\section{Conclusions}

We use the pfaffian combinatorial method to determinate the equilibrium properties of a system of adsorbed dimers in a 3-4 lattice. We consider four distinct types of bonds, with four values of energy. The results suggest that the specific heat per particle blows up when the fugacities associated to lattice bonds obey the relation $v = x + y$.

Although the 3-4 lattice showed in this paper it seems with the triangular lattice, our result demonstrates that the critical behavior is distinct for of that. In fact, in the present case appear a critical behavior nontrivial, which the critical temperature occurs up $T = 0$.

Much is already known about classical dimer modes in literature. Despite, as we show in this paper, some researchers still are interested in the treatment that models. In fact, this paper show that yet there are new interesting results for obtaining.




\begin{thebibliography}{50}
         
         \bibitem{fowler} R. Fowler and G. Rushbrooke, Trans. Faraday. Soc., \textbf{33} (1937) 1272.
         
         
         \bibitem{kasteleyn61} P. W. Kasteleyn, Physica \textbf{27} (1961) 1209.
         
         
         \bibitem{fisher61} M. E. Fisher, Phys. Rev. \textbf{124} (1961) 1664.
         
         
         \bibitem{fisher61_2} M. E. Fisher, J. Math. Phys. \textbf{7} (1961) 1776.
         	
         
         \bibitem{temperley61} H. N. V. Temperley, M.E. Fisher, Philos. Mag. \textbf{6} (1961) 1061.
         
         
         \bibitem{li} S. Li and W. Yan, Physica A \textbf{452} (2016) 251.
         
         
         \bibitem{giuliani} A. Giuliani, I Jauslin and E. Lieb, J. Stat. Phys. \textbf{163} (2016) 211.
         
         
         \bibitem{aizenman} M. Aizenman, M. Valcázar and S. Warzel, J. Stat. Phys. \textbf{166} (2017) 1078.
         
         
         \bibitem{casagrande} H. L. C. Casagrande, S. R. Salinas and F. A. da Costa. Braz. J. Phys. \textbf{41} (2011) 86.
         
         
         \bibitem{wu2016} R. Wu and W. Yan. Physica A. \textbf{457} (2016) 465.
         
         
         \bibitem{kasteleyn63} P.W. Kasteleyn, J. Math. Phys. \textbf{4} (1963) 287.
         
         
         \bibitem{nagle-yokoi} J. F. Nagle, C. S. O. Yokoi and S. M. Bhattacharjee, in \textit{Phase Transitions and Critical Phenomena}, ed. by C. Domb and J. Lebowitz (Academic Press, London, 1989), v. 13, p. 235 - 297.
         
         
          \bibitem{kenyon} H. Cohn, R. Kenyon, J. Propp, J. Amer. Math. Soc. \textbf{14} (2000) 297.
          
          \bibitem{montroll} E. W. Montroll, in \textit{Applied Combinatorial Methods}, ed. by E. F. Beckenback (Willey, New York, 1964), Chap. 4, p. 105 - 121.
          
          
          \bibitem{wu} F. Y. Wu, Int. J. Mod. Phys. B \textbf{20} 32 (2006) 5357.
          
          
          \bibitem{salinas-nagle} S. R. Salinas and J. F. Nagle. Phys. Rev. B \textbf{9} (1974) 4920.

                 
\end{thebibliography}
\end{document}